\documentclass[reprint,aps,twocolumn,superscriptaddress,preprintnumbers,letterpaper,natbib,floatfix,nofootinbib]{revtex4-2}

\RequirePackage{arydshln}
\usepackage{graphicx} % Required for inserting images

\makeatletter
\def\simgt{\mathrel{\lower2.5pt\vbox{\lineskip=0pt\baselineskip=0pt
           \hbox{$>$}\hbox{$\sim$}}}}
\def\simlt{\mathrel{\lower2.5pt\vbox{\lineskip=0pt\baselineskip=0pt
           \hbox{$<$}\hbox{$\sim$}}}}
\makeatother
\usepackage{latexsym}
\usepackage{braket}
\usepackage{amssymb}
\usepackage{epsfig,amsmath,graphics,tikz,tikz-feynman}
\usepackage{listings}
\usepackage{color}
\usepackage{dcolumn}
\usepackage{slashed}
\usepackage{cancel}
\usepackage{comment,latexsym} 
\usepackage{bm}
\usepackage{verbatim}

\usepackage{tabularx}
\usepackage{arydshln}

\usepackage{hhline}

\usepackage{dcolumn}
\definecolor{Mahogany}{rgb}{0.62,0.24,0.15}
\definecolor{colorLink}{rgb}{0.7,0,0}
\definecolor{colorCite}{rgb}{0,.7,0}
\definecolor{colorURL}{rgb}{0,0,0.7}
\usepackage[colorlinks=true,linktocpage=true,linkcolor=black,citecolor=colorCite,urlcolor=colorURL]{hyperref}
\usepackage{xspace}
\usepackage{multirow}
\usepackage{titlesec}
\usepackage[bbgreekl]{mathbbol}
\usepackage{bm}
\usepackage{float}
\usepackage{placeins}
\usepackage{afterpage}
\usepackage{footmisc}
\usepackage{breakcites}
\usepackage[most]{tcolorbox}
\usepackage{appendix,bm}
\tcbset{highlight math style={enhanced,colframe=green!30!black,colback=white,arc=5pt,boxrule=1.5pt}}

\usepackage{hyperref}
\usepackage{cleveref}

\newcommand{\bea}{\begin{eqnarray}}
\newcommand{\eea}{\end{eqnarray}}
\newcommand{\Y}{\mathcal{Y}}
\newcommand{\M}{\mathcal{M}}
\newcommand{\A}{\mathcal{A}}
\newcommand{\Mbar}{\bar{\mathcal{M}}}
\newcommand\psiB{\psi_{\mathcal{B}}}
\newcommand\psiBbar{\bar{\psi}_{\mathcal{B}}}
\newcommand{\Bsm}{\mathcal{B}_{\rm SM}}
\newcommand{\Bbarsm}{\bar{\mathcal{B}}_{\rm SM}}
\newcommand{\phiB}{\phi_{\mathcal{B}}}
\newcommand{\phiBbar}{\phi^*_{\mathcal{B}}}
\newcommand{\YBAU}{Y_{\mathcal{B}}^{\rm obs}}
\newcommand{\dcp}{\text{Dark}_{\rm CP}}

\begin{document}
\title{$\dcp$ Mesogenesis and New Implications for Collider Searches}

\author{Gilly Elor}
\email{gilly.elor@austin.utexas.edu}
\affiliation{Stone Aerospace, Del Valle, TX 78617}
\affiliation{Theory Group, The Weinberg Institute for Theoretical Physics, University of Texas at Austin, Austin, TX 78712, United States}

\begin{abstract}
We introduce $\dcp$ Mesogenesis in which  Standard Model mesons, either $B_{s,d}^0$, $B^\pm$ or $B_c^\pm$, undergo out-of-equilibrium dark CP violating decays to  Standard Model and dark sector baryons. With order one CP violation in the dark sector, the observed baryon asymmetry of the Universe can then be generated with branching fractions as small as $2.7 \times 10^{-8}$ for the exotic decay of $B$-mesons into baryons and missing energy. This sets a lower limit on the branching fraction below which Mesogenesis becomes disfavored. 
%current experimental limits on this observable are at the $10^{-6}$ level; one to two order of magnitude improvement should be possible with existing colliders.  
This work therefore strongly encourages collider searches to target sensitivities one to three orders of magnitude below planned limits. For completeness, we further introduce a variant on $\dcp$ Mesogenesis involving dark lepton states. We conclude by laying out a roadmap for the complete exclusion (or discovery) of the Mesogenesis framework. 
\end{abstract}

\maketitle
 \tableofcontents 
\allowdisplaybreaks

%%%%%%%%%%%%%%%%%%
\section{Introduction}
\label{sec:intro}
%%%%%%%%%%%%%%%%%%
The existence of the Universe as we know it requires a mechanism for generating a small asymmetry of matter over anti-matter at early times. 
From measurements of the Cosmic Microwave Background (CMB) \cite{Ade:2015xua,Aghanim:2018eyx} and light element abundances  \cite{Cyburt:2015mya,pdg}, the value of the baryon asymmetry of the universe (BAU) has been determined, in co-moving yield units, to be   
$\YBAU \equiv (n_{\mathcal{B}}-n_{\bar{\mathcal{B}}})/s = \left( 8.718 \pm 0.004 \right) \times 10^{-11}$. The nature of the mechanism which generated this small asymmetry remains an outstanding mystery which the Standard Model of Particle Physics (SM) fails to explain. 

Since the 1960s, many baryogenesis mechanisms have been proposed which satisfy the three requisite Sakharov conditions~\cite{sakharov} for generating the BAU; \emph{i.} baryon number violation, \emph{ii.} C and CP violation, and \emph{iii.} interactions which happen out-of-thermal equilibrium. However, the experimental verification of the nature of almost all baryogenesis proposals remains challenging if not impossible \cite{Elor:2022hpa}---gone it seemed were the days where the scientific admissibility of a proposed theory was guided by the criteria that it be observationally refutable \cite{popper}.   However, in recent years a new class of baryogenesis mechanisms have been proposed;  mechanisms of Mesogenesis \cite{Elor:2018twp,Elor:2020tkc,Elahi:2021jia,Elor:2024cea} are testable and predict signals at current experiments. In this work we introduce a new (and arguably the simplest) variant of Mesogenesis. We then tackle the question of whether Mesogenesis (in any variant) can ever be fully excluded?

Mesogenesis mechanisms \cite{Elor:2018twp,Elor:2020tkc,Elahi:2021jia,Elor:2024cea} generate the BAU, along with the dark matter relic abundance, using out-of-equilibrium processes already present in the SM---  decays of heavy mesons. Thus, by construction, mechanisms of Mesogenesis must generate the BAU at late times; after QCD confinement but before the epoch of Big Bang Nucleosynthesis (BBN). In Mesogenesis ``baryon number violation"  is satisfied by letting dark states carry Standard Model baryon number, $Q_\mathcal{B}$, such that equal and opposite amounts of baryons can be sequestered via meson decays into the Standard Model baryons and the dark matter. 

Generating $\YBAU$ in existing Mesogenesis proposals \cite{Elor:2018twp,Elor:2020tkc,Elahi:2021jia,Elor:2024cea} requires that the product of experimental observables be roughly $\left( A_{CP}^{\rm visible} \times \prod \text{Br} \right) \gtrsim 10^{-5}$, where $A_{CP}^{\rm visible}$ parametrizes the CP visible sector violation in the meson system (e.g. the semi-leptonic asymmetry in neutral $B$ oscillations \cite{Elor:2018twp}) and $\prod \text{Br}$ is taken to be the product of branching fractions for every decay involved in the process\footnote{In Mesogenesis variations involving dark leptons \cite{Elor:2020tkc,Elahi:2021jia} the dark sector is populated by a cascade of decays.}. The novelty of Mesogenesis lies in its testability e.g., through improved measurements of $A_{CP}^{\rm visible}$ (e.g., at Belle-II \cite{Zlebcik:2025ubm}), and dedicated $B$-Mesogenesis searches \cite{Belle:2021gmc, BaBar:2023rer,BaBar:2023dtq,BaBar:2024qqx} which have set limits on on the branching fraction of exotic $B$ meson decays. 

\begin{figure}[t!]
\centering
\includegraphics[width=0.5\textwidth]{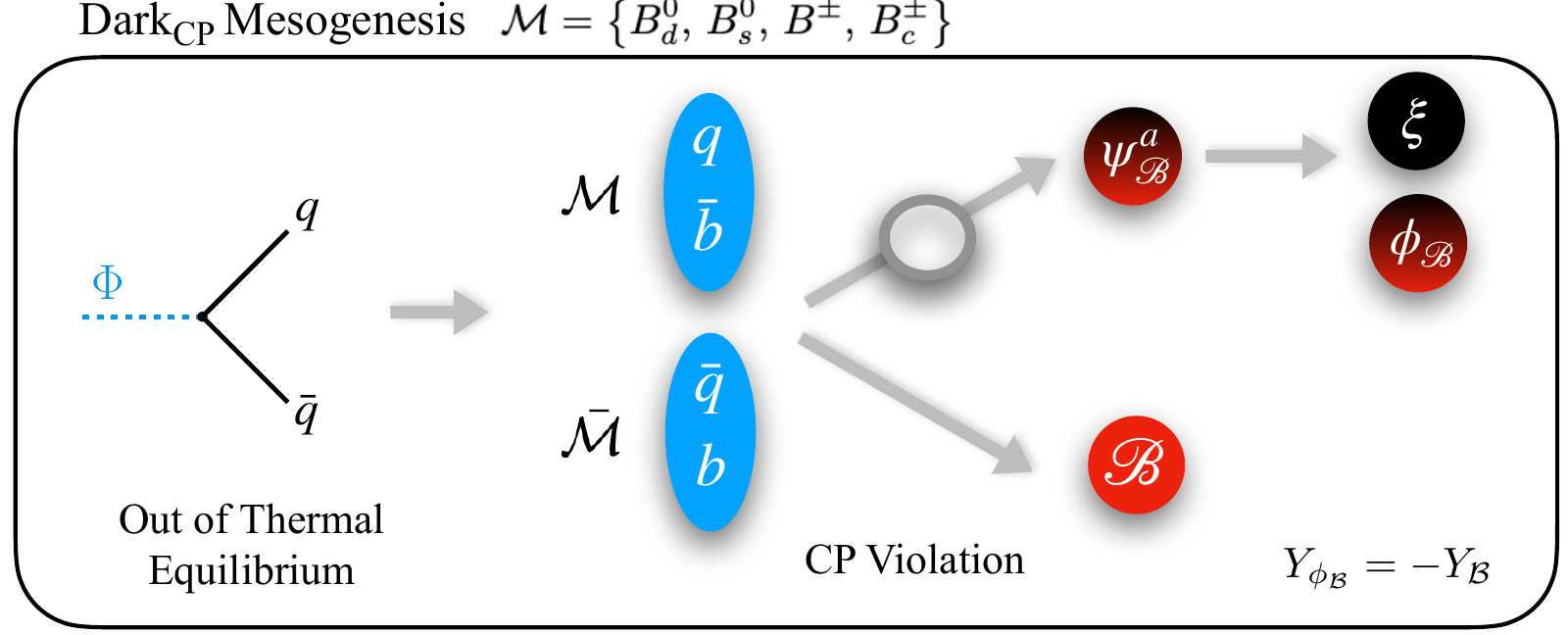}
\vspace{-.2cm}
\caption{A depiction of the $\dcp$ Mesogenesis mechanism where the mesons $\M$ undergo kinematically allowed decays to Standard Model and dark baryons.  CP violation arises in the $\M$ meson decay itself through 1-loop and tree level interference from diagrams with additional dark sector fermion flavors. }
\label{fig:cartoonBaryon}
\end{figure}

In this paper we introduce $\dcp$ Mesogenesis: a variant of Mesogenesis in which the CP violation arises entirely from a dark sector while retaining testability of the mechanism. The BAU can be generated through exotic $B^0_{s,d}$, $B^\pm$, or $B_c^\pm$ meson decay to SM baryons $\mathcal{B}$ and missing energy. Like other variants of Mesogenesis, dark sector states carry baryon (or lepton) number and are viable dark matter candidates. SM meson decays, combined with CP violating phases in the dark sector sequester equal and opposite amounts of baryon number in the two sectors, generating the BAU as well as the dark matter relic abundance. $\dcp$ Mesogenesis is arguably the simplest variant proposed to date, as there is no need for particle oscillations \cite{Elor:2018twp} or cascade decays \cite{Elor:2020tkc,Elahi:2021jia}. Figure~\ref{fig:cartoonBaryon} depicts the $\dcp$ Mesogenesis mechanism. 

From a completely model agnostic standpoint, the dark CP-violation can be $\mathcal{O}(1)$ while remaining consistent with all existing experimental observations. 
Consequently, as we will show, the BAU can be generated with branching fractions as small as $10^{-8}$. This represents a bound on the smallest possible $B \rightarrow \mathcal{B} + \text{missing energy}$ through which any variant of Mesogenesis (involving dark baryons) can occur. This sets a lower limit on the branching fraction below which Mesogenesis becomes disfavored. As such, this work provides a benchmarks to guide experimental searches towards the discovery (or ultimately the exclusion) of the Mesogenesis framework \footnote{Variations of Mesogenesis involving a mass shift in the early Universe \cite{Elor:2024cea, technion} are exceptions to this statement as the present day signal is ``turned off". In \cite{Elor:2024cea} this required a very fine-tuned dark sector phase transition. Upcoming work will explore the degree to which tuning in this context can be relaxed \cite{technion}.}. 

%This paper is organized as follows: In Sec. etc. \GE{Fill in}

\newpage
%%%%%%%%%%%%%%%%%%
\section{$\dcp$ $\M$-Mesogenesis} 
\label{sec:mech}

We now introduce the $\dcp$ Mesogenesis mechanism. We derive the corresponding Boltzmann equations and solve them to explore the mechanism parameter space where the observed BAU is generated. We remain as agnostic as possible about the dark sector dynamics. 

\begin{table}[t!]
\renewcommand{\arraystretch}{1.25}
  \setlength{\arrayrulewidth}{.25mm}
\centering
\small
\setlength{\tabcolsep}{0.18 em}
\begin{tabular}{|c || c | c | c | c  | }
\hhline{- - - - -}
   $\mathcal{M}$ &   $\mathcal{O}_{bud}$   &  $\mathcal{O}_{bus}$ &  $\mathcal{O}_{bcd}$ & $\mathcal{O}_{bcs}$      \\ 
\hhline{- - - - -}
$B_d^0$  &   $n \, (udd)$   &  $\Lambda^0 \, (uds)$ &  $\Sigma_c^0 \, (cdd)$ &  $\Xi_c^0 \, (dcs)$      \\ 
$\Delta M \, [\text{MeV}]$ & $4340.1 $ & $4164.0 $ & $2825.9$ & $2808.8 $ \\
\hhline{- - - - -}
$B_s^0$  & $\Lambda^0 \, (uds)$ & $\Xi^0 \, (uss)$ & $\Xi_c^0 \, (dcs)$ & $\Omega_c^0 \, (ssc)$ \\
 $\Delta M \, [\text{MeV}]$ & $4251.2 $ & $4052.0$ & $2896.0 $  & $2671.7$ \\
\hhline{- - - - -}
$B^+$  & $p \, (uud)$ & $\Sigma^+ \, (uus)$ & $\Lambda_c^+ \, (udc)$ & $\Xi_c^+ \, (usc)$ \\
$\Delta M \, [\text{MeV}]$ & $4341.1$ & $4090.0 $ & $2992.9 $ & $2811.4\, \text{MeV}$ \\
\hhline{- - - - -}
$B_c^+$  & $\Lambda_c^+ \, (cud)$ & $\Xi_c^+ \, (cus)$ & $\Xi_{cc}^+ \, (dcc)$ & $\Omega_{cc}^+ \, (scc)$ \\
$\Delta M \, [\text{MeV}]$ & $3988.4$ & $3807.0$ & \emph{unknown} & \emph{unknown} \\
\hhline{- - - - -}
\end{tabular}
\caption{For each $\mathcal{M}$ we list the daughter SM baryon arising from the decay of the $\bar{b}$ via the four possible flavorful operators, $\mathcal{O}_{b,u_i d_j} \equiv i \epsilon_{\alpha \beta \gamma} b^\alpha (\bar{d}_j^{c \beta} u_i^\gamma)$, resulting in the decay  $\mathcal{M} \rightarrow \mathcal{B}_{\rm SM} \, \psiB$. The energy released is $\Delta M \equiv M_{\mathcal{M}} - M_{\mathcal{B}_{\rm SM}}$. 
}
\label{tab:decays}
\end{table}

%%%%
\subsection{The Mechanism}
\label{sec:mech}
%%%
As with all variants of Mesogenesis, the starting point for $\dcp$ Mesogenesis is the assumption of a late matter dominated era where the energy density of the Universe is dominated by a scalar field $\Phi$ with mass $m_\Phi \sim \mathcal{O}(10-100)$~GeV. It is then further assumed that $\Phi$ decays to quarks $\Phi \rightarrow q \bar{q}$, injecting energy into the plasma when the temperature of the Universe was of order $T_R \sim 5-80$ MeV i.e. $\Gamma_\Phi \sim H(T_R) \sim 10^{-21} \text{GeV}^{-1}$. At these scales---before the era of BBN but well below the QCD phase transition--- the produced quarks will immediately hadronize to form charged and/or neutral Standard Model mesons $\M$.  We assume a symmetric reheating with no CP violation in $\Phi$ decays; mesons and anti-mesons are produced at equal rates and we define the branching fraction $\text{Br}^{\M}_\Phi \equiv \text{Br} (\Phi \rightarrow \M) = \text{Br} (\Phi \rightarrow \Mbar)$ to implicitly incorporate fragmentation functions.

In analogy to visible sector CP violating Mesogenesis constructions, we assume the existence of baryon number preserving interaction between Standard Model quarks and a dark sector, where dark sector states are charged under Standard Model baryon number.  We introduce a heavy colored scalar mediator $\Y$, and three or more Standard Model singlet Dirac fermions $\psiB^{\alpha}$, with a general mass matrix
$\mathcal{L}_{mass}  = - \sum_{\alpha \beta} M_{\alpha \beta} \psiBbar^\alpha \psiB^\beta$, where $\alpha, \beta$ are dark flavor indices.  
In the mass basis for the $\psiB^\alpha$s we take $\psiB^1$ to be the lightest state, $\psiB^2$ to be the next lightest etc. $\Y$ and the $\psiB$ carry baryon number, and the following interactions with the Standard Model are allowed:
\bea
\hspace{-0.18in} \mathcal{L}_{\Y_{2/3}}  = \,\, && - \sum_{i,j}  \tilde{y}_{ij} \, \epsilon_{abc} \, \Y^{a} d^b_{Ri} d^c_{Rj} \\ \nonumber
&& \qquad - \sum_{k,\alpha} y_{k\alpha} \Y^*_a u_{Rk}^a \psi^\alpha_{\mathcal{B}, R} + \text{h.c}\,,
\label{eq:L23}
\eea
where $a, b, c$ are color indices and $R$ denotes the right-chiral components of each field (only the $SU(2)_{W}$ Standard Model singlet fields couple to the dark sector). Note that with the quantum numbers\footnote{Alternative charge assignments are also possible ~\cite{Alonso-Alvarez:2021oaj}.} assignments presented in Table~\ref{tab:newparticles}, $\psiB$s cannot mix with the Standard Model quarks. The mass of $\Y$ is constrained to be roughly above the TeV scale by collider searches \cite{Alonso-Alvarez:2021qfd}. 

Far below the mass of $\Y$, the interaction between the two sectors can be written as the four-fermion operator: 
\bea
\mathcal{O}^l_{d_i u_j d_k} = \frac{\tilde{y}_{ij} y_{k\alpha}}{M_{\Y}^2} \epsilon_{abc} d^b_{Ri} d^c_{Rj} u_{Rk}^a \psi^\alpha_{\mathcal{B}, R}\,,
\label{eq:op}
\eea
which mediates the out-of-equilibrium decay Standard Model mesons $\M$ into $\psiB$s and a Standard Model baryon.  Since $\psiB$s carry baryon number, their mass must lie in the range: 
\bea 
m_{\M}- m_{\Bsm} > m_{\psiB} > m_p + m_e  = 938.78 \, \text{MeV}\,. 
\label{eq:kin}
\eea 
This condition kinematically forbids proton decay---attempts to suppress proton decay through a judicious choice of flavor structure do not succeed since loop level contributions involving quark mixing are still above the relevant experimental bounds. From the requirement in Eq.~\eqref{eq:kin}, it is evident that the operator in Eq.~\eqref{eq:op} can lead to the kinematically allowed decay of the following mesons: 
\bea
\M = \left\{ B_d^0, \, B_s^0, \, B^\pm, \, B_c^\pm \right\}\,.
\eea 

In $\dcp$ Mesogenesis, the CP violating decay of the Standard Model mesons,   
\bea
\Gamma \left( \M \rightarrow \psiB^\alpha \, \Bsm \right) \neq \Gamma \left( \Mbar \rightarrow \psiBbar^\alpha  \, \Bbarsm \right)\,,
\eea
arises from the interactions in the dark sector thus satisfying the C and CP violation Sakharov conditions. As the mesons (and anti-mesons) decay out of equilibrium, baryon number can thus be transferred to the dark sector, leading to an ``apparent violation'' of baryon number in the visible sector. Consequently, equal and opposite baryon number abundances will be generated in the dark and visible sectors. Figure ~\ref{fig:cartoonBaryon} illustrates the $\dcp$ Mesogenesis process. Table~\ref{tab:decays} lists all kinematically allowed final state baryons. These arise from the parton level $\bar{b}$ quark decays\footnote{Note that in $\dcp$ $B_c^+$ Mesogenesis, any $c$ quark decays, $c \rightarrow \bar{d}_i \bar{d}_j \psiBbar$, would counteracting the baryon asymmetry generation from the $\bar{b}$ decay. Thankfully these are all kinematically forbidden by the expected mass of the final state beautiful baryon.} mediated through Eq.~\ref{eq:op}. 

To avoid washing out the produced asymmetries, the lightest $\psiB$ must predominantly decay to dark matter particles, stabilized by an additional $\mathbb{Z}_2$. We therefore must also minimally introduce a charged scalar $\phiB$ carrying baryon number (with GeV scale mass to ensure the stability of matter), and a dark Majorana fermion $\xi$, with the $\mathbb{Z}_2$ invariant (see charge assignments in Table~\ref{tab:newparticles}) interaction $\mathcal{L}_{dark} = - y_{dark, \alpha}  \psiB^\alpha \phiB \xi  + \text{h.c.}, $ where at the very least $y_{dark, \alpha}  \neq 0$ so that the lightest state $\psiB^1$ quickly decays $\psiB^1 \rightarrow \phiB \xi$ within the dark sector $\phiB$ and $\xi$ are both stable particles contributing to the dark matter abundance. The observed BAU is therefore related to the dark sector abundances of stable particles: $-Y_{\mathcal{B}} =Y_{\phiB - \phiBbar}=  \left( n_{\phiB} - n_{\phiBbar} \right) /s$. Note that the dark matter production, constraints, and signals are identical to every other variant of Mesogenesis involving dark sector baryons---for a review see \cite{Elor:2022hpa} and see Figure 4 of \cite{Berger:2023ccd} for the $(m_\xi, m_{\phiB})$ parameter space. 

%%%%%%%%%%%%%
\subsection{Generating the Baryon Asymmetry}
\label{subsec:BA}
%%%%%%%%%%%%
As discussed above we assume a late matter dominated era $\rho_{rad} \ll m_\Phi n_\Phi$. In this era, the Boltzmann Equations for radiation and $\Phi$ are given by:
\begin{subequations}
\begin{align}
 \frac{d n_\Phi}{dt} + 3 H n_\Phi &= - \Gamma_\Phi n_\Phi \,, \\ 
  \frac{d \rho_{\rm rad}}{dt} + 4 H \rho_{\rm rad} &= \Gamma_\Phi m_\Phi n_\Phi \,,\\
   H^2 &= \frac{8\pi}{3 M_{\rm pl}} \left(\rho_{\rm rad} + m_\Phi n_\Phi \right), 
\end{align}
\label{eq:PhiRadBE}
\end{subequations}
%We remain agnostic about the nature of $\Phi$ and only assume that it initially dominates the energy density of the Universe. 

For a choice of meson $\M$, the evolution of meson number density is governed by: 
\bea
 \frac{d n_\M}{dt} + 3 H n_{\M} = \Gamma_\Phi \text{Br}_{\Phi}^{\M} n_\Phi - \Gamma^\M_{\rm tot} n_{\M}\,, %\\ 
%& \frac{d n_{\Mbar}}{dt} + 3 H n_{\Mbar} = \Gamma_\Phi \text{Br}_{\Phi }^{\M} n_\Phi - \Gamma^\M_{\rm tot} n_{\Mbar},
\label{eq:MesonBE}
\eea
and similarly for anti-mesons.  We have taken the total decay width of $\M$ to be $\Gamma_{\M} \sim \Gamma_{\Mbar}$, as the branching fraction of the new CP violating decay modes will be small. Since $H \ll \Gamma_\Phi \ll \Gamma_{\M}$ in the era of interest, we can neglect Hubble expansion and assume rapid meson decays \cite{Elor:2020tkc,Elahi:2021jia}. With this assumption  Eq.~\eqref{eq:MesonBE} simply reduces to $
 \Gamma^\M_{\rm tot} n_{\M} \simeq  \Gamma^\M_{\rm tot} n_{\Mbar}  \simeq  \Gamma_\Phi \text{Br}_{\Phi}^{\M} n_\Phi$. 
The abundances of dark baryons $\psiB$s is then governed by: 
\bea
 \frac{d n_{\psiB^\alpha}}{dt} + 3 H n_{\psiB^\alpha} =  \Gamma^{\M}_{\rm tot} \text{Br}_{\mathcal{M}}^\alpha  n_{\M}  \simeq  \Gamma_\Phi \text{Br}_{\Phi}^{\M}  \text{Br}_{\mathcal{M}}^\alpha  n_\Phi \,, 
 \label{eq:psiBBE}
\eea
where we have defined $\text{Br}_{\mathcal{M}}^\alpha \equiv \text{Br}( \M \rightarrow \Bsm + \psiB^\alpha)$.  An analogous expression tracks the abundance of anti dark baryons $\psiBbar^\alpha$. 
All particles have GeV-scale masses and are far from thermal equilibrium at MeV temperatures, so we omit equilibrium number density terms from the Boltzmann evolution. 

\newpage
Since at least one of the $\psiB$s quickly decays into dark matter states $\phiB$ and $\xi$, the SM baryon asymmetry evolves as:
\bea
 \frac{d (n_{\phiBbar} - n_{\phiB})}{dt}+ &&  3 H (n_{\phiBbar} - n_{\phiB}) \\ \nonumber
&&  = \, \Gamma_\Phi \text{Br}_\Phi^{\M} n_\Phi \sum_\alpha \left(
\text{Br}_{\Mbar}^\alpha  - \text{Br}_{\M}^\alpha \right) \\ \nonumber
&& \simeq \, 2\, \Gamma_\Phi \text{Br}_\Phi^{\M} n_\Phi 
\, \sum_\alpha A_{CP}^{ dark, \alpha} 
\text{Br}_{\M}^\alpha \,,
\label{eq:asymBE}
\eea
where the \emph{dark CP asymmetry} $A_{CP}^{dark,\alpha}$ is defined  in analogy to the visible sector CP violation:
\bea
A_{CP}^{dark,\alpha} &\equiv &
 \frac{
 \Gamma (\Mbar\rightarrow \psiBbar^\alpha \Bbarsm) - \Gamma (\M \rightarrow \psiB^\alpha \Bsm)}{ \Gamma (\Mbar\rightarrow \psiBbar^\alpha \Bbarsm) + \Gamma (\M \rightarrow \psiB^\alpha \Bsm)}.
 \label{eq:ACPdark}
\eea
Note that we have chosen the baryon number of $\psiB$ and $\phiB$ such that positive $A_{CP}^{\rm dark,\alpha}$ generates the BAU. However $A_{CP}^{\rm dark, \alpha} < 0$ can  successfully  generate the BAU by simply swapping the baryon number of $\psiB$ and $\psiBbar$. For generality, in this section we have remained as agnostic as possible about the dark sector model and simply observe from Eq.~\eqref{eq:ACPdark} that the generated visible baryon asymmetry, $Y_{\mathcal{B}} = -(n_{\phiBbar} - n_{\phiB})/s$, will be proportional to $\sum_\alpha \text{Br}_\M^\alpha |A_{CP}^{\rm dark,\alpha}| \equiv \text{Br} \times |A_{CP}^{\rm dark}|$, where $\alpha$ sums over different flavors of dark fermionic baryons that contribute to the production of $\phiB$ and $\xi$. 

\begin{figure}[t!]
\centering
\vspace{0.2in}
\includegraphics[width=0.5\textwidth]{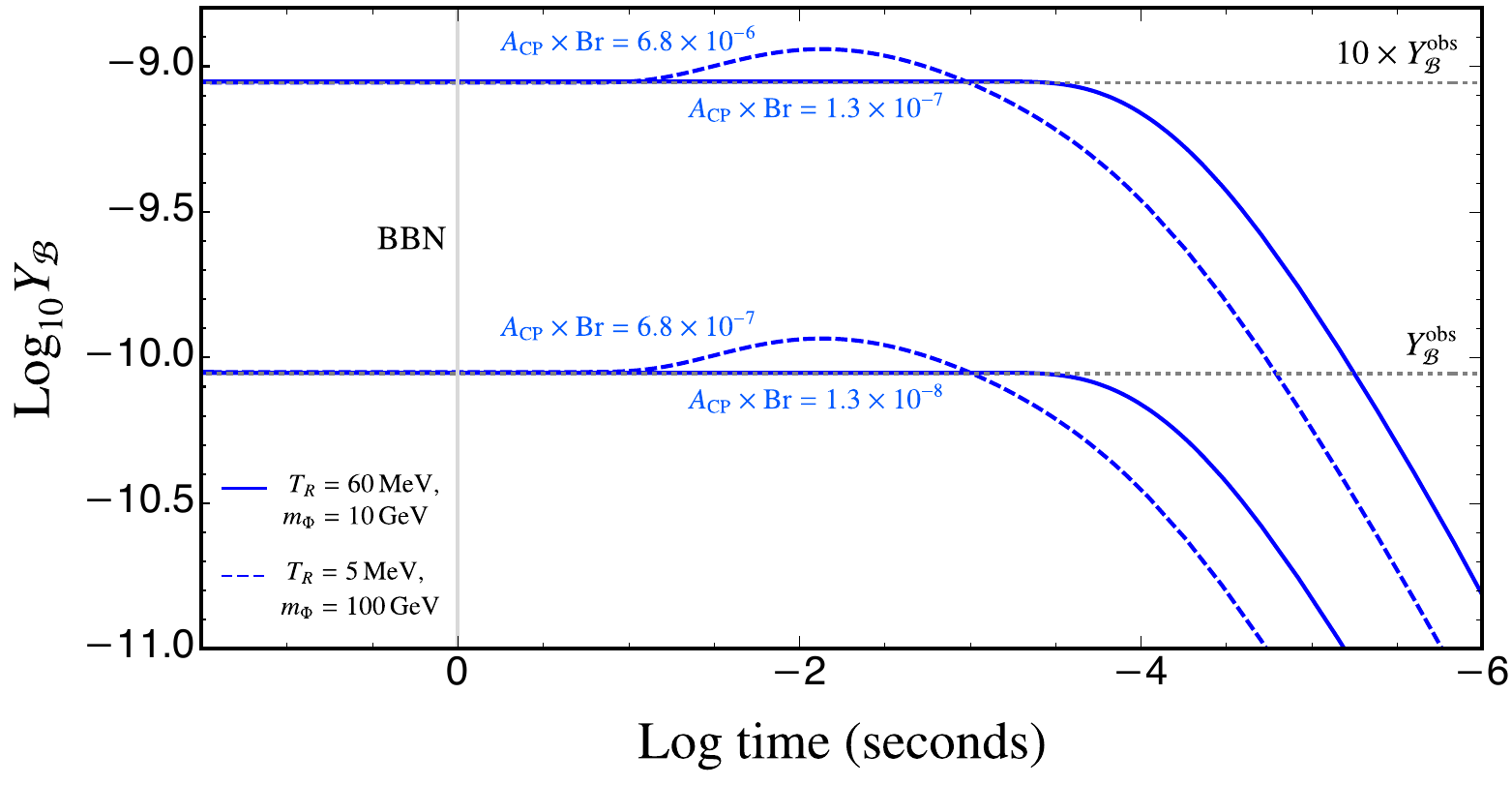}
\vspace{-.2cm}
\caption{Benchmark points corresponding to solving the Boltzmann equations for the baryon asymmetry Eq.~\eqref{eq:asymBE} in conjunction with Eq.~\eqref{eq:PhiRadBE}. $\text{Br}_\Phi^{\M}$ is assumed to be one for a given $\M$ under consideration. Note that the small differences between dashed and dotted line implies results will be relatively insensitive to $T_R$. 
}
\label{fig:BAUyeild}
\vspace{-0.2in}
\end{figure}

\vspace{0.03 in}
\subsection{Parameter Space}
Eq.~\eqref{eq:asymBE} can be numerically integrated in conjunction with Eq.~\eqref{eq:PhiRadBE} following the prescription outlined in \cite{Elor:2018twp,Elor:2020tkc}. 
Figure~\ref{fig:BAUyeild} shows the resulting visible sector baryon yield, $Y_{\mathcal{B}}$, evolution for four benchmark values of $\{T_R, m_\Phi,  \text{Br}_{\M}  \times A_{CP}^{\rm dark}\}$. 
Eq.~\eqref{eq:asymBE} was derived assuming $\M$ decays dominated the Boltzmann equations over possible $\M$ annihilation---where typical cross sections for hadrons are $\sigma \simeq m_\pi^{-1} \simeq \mathcal{O}(10\text{mb})$. Following the approach in \cite{Elor:2020tkc,Elahi:2021jia}, for all $\M$ we consider here i.e. with lifetimes of $ \{ \tau_{B_d^0}, \tau_{B_s^0}, \tau_{B^+}, \tau_{B_c^+}\}  = \left\{ 1.52,\,1.51,\, 1.64 \,, 0.51 \right\}  \times 10^{-12}$ seconds, we roughly expect (ignoring scatterings) Eq.~\eqref{eq:asymBE} to be valid up to $T_R^{\rm max} = 60 \, \text{MeV}$.  $\dcp$ Mesogenesis can thus produce the BAU over a wider range of $T_R$ values in comparison to visible sector CP violation Mesogenesis variants \cite{Elor:2018twp,Elor:2020tkc,Elahi:2021jia,Elor:2024cea} where additional processes prevent or reduce the BAU generation at higher values of $T_R$ (e.g., decoherence of $B^0$ oscillations \cite{Elor:2018twp}).

Benchmarks in Figure~\ref{fig:BAUyeild} are chosen to represent minimum and maximum CP violation and branching fraction required to generate the BAU. In particular,  $(T_R^{\rm min}, m_\Phi^{\rm max}) = (5 \,\text{MeV}, 100\, \text{GeV})$ requires $A_{CP}^{\rm dark} \times \text{Br}_{\M} = 6.8 \times 10^{-7}$ to generate $\YBAU$. By contrast, for $(T_R^{\rm max}, m_\Phi^{\rm min}) = (60\, \text{MeV}, 10\, \text{GeV})$ the product $A_{CP}^{\rm dark} \times \text{Br}_{\M}$ can be as low as $1.3 \times 10^{-8}$. Also shown in Figure~\ref{fig:BAUyeild} is the ``washout regime" where the baryon asymmetry is initially over-produced $Y_{\mathcal{B}} > \YBAU$ (for instance due to higher temperature effects). We leave the numerical analysis of $\dcp$ Mesogenesis in the washout regime to future work \cite{washout}. 

\begin{figure}[t!]
\centering
%\vspace{-0.05in}
\includegraphics[width=0.4\textwidth]{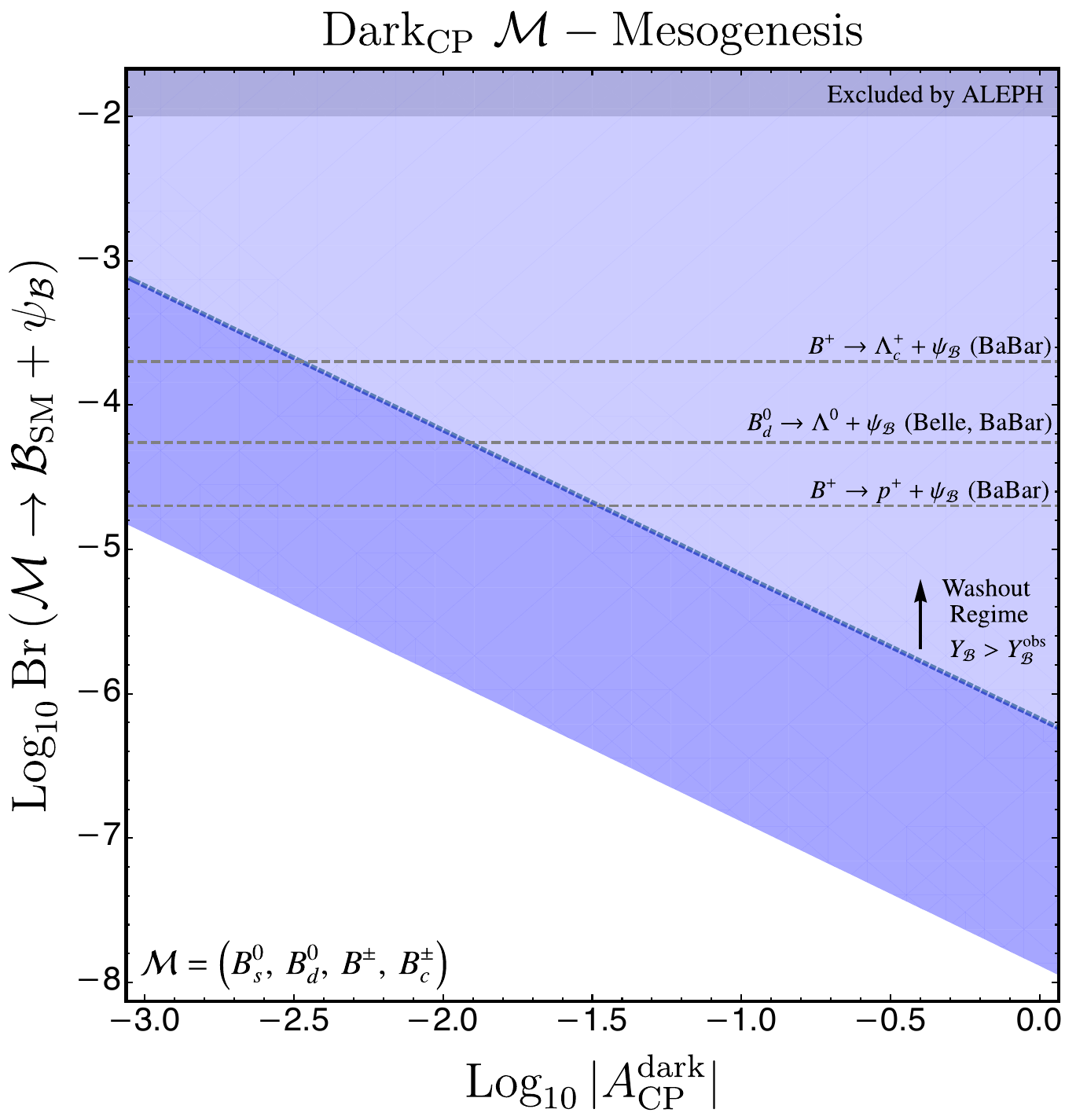}
%\vspace{-0.05in}
\caption{ Parameter space for successful $\dcp$ Mesogenesis is shown in blue. Light blue corresponds to the washout regime where the baryon asymmetry is initially overproduced. Dotted lines correspond to existing neutral $B$ Mesogenesis searches \cite{BaBar:2023rer,Belle:2021gmc,BaBar:2023dtq,BaBar:2024qqx}, while the branching fractions greater than $10^{-2}$ are excluded by ALEPH \cite{Alonso-Alvarez:2021qfd}.}
\vspace{-0.2in}
\label{fig:param}
\end{figure}

In Figure~\ref{fig:param} we display the full mechanism parameter space---where $\dcp$ Mesogenesis successfully generates the BAU. Numerically integrating the Boltzmann equations: 
\bea
\hspace{-0.25in}
\boxed{
\frac{Y_{\mathcal{B}}}{Y_{\mathcal{B}}^{obs}} \simeq \frac{  \text{Br} }{1.3 \times 10^{-8}}  \bigl|A_{CP}^{dark} \bigr|\frac{T_R}{60 \, \text{MeV}} 
\frac{2m_{\M}}{m_\Phi}\, ,
}
\label{eq:BAU}
\eea
where the branching fraction for $\M$ decays to Standard Model baryons and missing energy is inclusive; over all final state baryons. 
%, and $\text{Br} \times |A_{CP}^{\rm dark}| \equiv \sum_i \text{Br}_\M^i |A_{CP}^{\rm dark,i}|$ \GE{if more than one decay contributes to the asymmetry}. 
The blue region in Figure~\ref{fig:param} corresponds to $T_R = 5-60\, \text{MeV}$. In the light-blue region the BAU is overproduced $Y_{\mathcal{B}} > Y_{\mathcal{B}}^{obs}$ corresponding to the washout regime where, for instance, efficient $\M$-$\bar{\M}$ annihilations above 60 MeV decrease the asymmetry (more exotic effects will be explored in upcoming work \cite{technion,washout}). Also shown are limits from existing designated Mesogenesis searches: $B_d^0 \rightarrow \Lambda^0 + \text{MET}$ by the Belle-II collaboration (using Belle-I data) \cite{Belle:2021gmc} $\&$ BaBar \cite{BaBar:2023rer}, BaBar searches for $B^+ \rightarrow p + \text{MET}$ \cite{BaBar:2023dtq}, and $B^+ \rightarrow \Lambda_c^+ + \text{MET}$ \cite{BaBar:2024qqx}. Recast limits from ALEPH \cite{Alonso-Alvarez:2021qfd} are shown in gray. These limits are well above the majority of $\dcp$ Mesogenesis parameter space. Furthermore, the smallest possible branching fraction $\text{Br} \times A_{CP}^{dark} \gtrsim 1.3 \times 10^{-8}$, which corresponding to $A_{CP} \sim \mathcal{O}(1)$, is below (but not far) the current targeted sensitivity of Belle-II~\cite{private}.

%%%%%%%%%%%%%%%%%%%
\section{New Implications for Collider Searches}
\label{sec:newimplications}
%%%%%%%%%%%%%%%%%%%
By entirely sequestering the CP violation into the dark sector, $\dcp$ Mesogenesis represents a general lower limit on the size of the branching fraction of the exotic $B$ meson decays into Standard Model baryons and missing energy such that the BAU can be generated. Critically, from Eq.~\eqref{eq:BAU}, we see that the maximum possible dark sector CP violation, i.e. order one phases yielding values of $A_{CP}^{\rm dark}$ of up to unity, corresponds to a lower limit on the branching fraction of:  
\bea
\boxed{
\text{Br} \left(\mathcal{M}\rightarrow \Bsm + \text{MET} \right) \gtrsim 10^{-8}.
}
\label{eq:masterbound}
\eea
Below this limit, our work shows that any variant of Mesogeneisis with dark baryons is no longer well motivated. On the flip side a measurement of $\text{Br}_{\M}$ in the range of $10^{-8}$-$10^{-5}$, while excluding visible sector CP violation variants of Mesogenesis, would be a signal of $\dcp$ Mesogenesis. Coupled with complementary probes and indirect signals $\dcp$ Mesogenesis could still be entirely reconstructible. 

\vspace{0.1in}
\noindent \emph{This work therefore strongly encourages experimentalists to adopt a target sensitivity on the branching fraction, of charged and neutral $B$ meson to baryons and missing energy,  at the level of $10^{-8}$ to $10^{-7}$}.

%%%%%%%%%%%%%%%%%%%
\subsection{Direct Signals at $B$-factories}
\label{sec:directSignals}
%%%%%%%%%%%%%%%%%%%

%Since the branching fraction of the exotic decay of $B$ mesons into SM baryons and missing energy is directly related to the BAU, the decay, would provide a powerful probe of $\dcp$ Mesogenesis. 

$B$-factories are extremely well suited to directly probe $\dcp \, B^\pm$ $\&$ $\dcp \, B^0_{s,d}-$Mesogenesis as they can directly search for the decays listed in  Table~\ref{tab:decays} that generate the BAU. Meanwhile, LHCb can probe the decays through which $\dcp$ $B_c^+$-Mesogenesis may proceeds, however, future colliders would likely be needed to fully constrain the parameter space of this variant. 

To date, several designated Mesogenesis searches have been conducted: Belle-I and BaBar data was used to constrain the branching fraction of $B_d^0 \rightarrow \Lambda_0 \psiB^1$ \cite{Belle:2021gmc, BaBar:2023rer} to a range of $10^{-5}$ to $10^{-4}$ over the kinematically allowed range of $m_{\psiB}$. BaBar data was additionally used to set similar limits on $B^+ \rightarrow p \psiB$ \cite{BaBar:2023dtq} and $B^+ \rightarrow \Lambda_c^+ \psiB$ \cite{BaBar:2024qqx}. While these searches begin to disfavor certain channels through which the original Neutral $B$-Mesogenesis mechanism \cite{Elor:2018twp} can proceed, there are still 3-4 allowed orders of magnitude in sensitivity through which $\dcp$ $B_{d,s}^0$ and $\dcp$ $B^+$ Mesogenesis can generate the BAU. 

Belle-II should have the capabilities to reach sensitivities down to the $10^{-7}$ level (possibly better) \cite{private}, approaching the lower limit on the branching fraction for which $\dcp$ $B^0_{s,d}$ and $\dcp \, B^+$ Mesogenesis can generate the entire BAU. Furthermore, given a detection, Belle-II should also be able to reconstruct the conjugate decay and therefore  measure $A_{CP}^{\rm dark}$ \cite{private}. Thus, both observables related to the BAU can be determined at Belle-II. On the flip-side, current $B$ factory Mesogenesis searches \cite{Belle:2021gmc, BaBar:2023rer} trigger on a final-state $\Lambda$ baryon (ignoring the conjugate channel). As such $A_{\rm CP}^{\rm dark}$ is currently unconstrained by direct collider searches (the strongest indirect constraints come from electric dipole moments as will be discussed below). 

%%%%%%%%%%%%%%%%%%%
\subsection{Indirect Probes at LHCb and Future Colliders}
\label{sec:HadronCollSignals}
%%%%%%%%%%%%%%%%%%%
Hadron colliders, and LHCb in particular, can also probe $\dcp$ Mesogenesis by searching for apparent baryon number violation in heavy baryon decays to mesons and missing energy~\cite{Rodriguez:2021urv}. Table~\ref{tab:bBaryonDecays} comprehensively (for the first time) lists all such possible decays which are kinematically allowed i.e. where $\Delta M > m_{\psiB} \gtrsim 1 \, \text{GeV}$. 

\begin{table*}[t]
\renewcommand{\arraystretch}{1.05}
  \setlength{\arrayrulewidth}{.2mm}
\centering
\small
\setlength{\tabcolsep}{0.18 em}
\begin{tabular}{ c  c }
\begin{tabular}{ |c || c | c | c  |}
\hhline{- - - -}
    Operator 			&  \,\, Initial \,\,  &  Final 	&     \,\,\, $\Delta M$     \,\,\,  \\     
    and Decay   			&  Baryon &  Mesons				& (MeV)       \\ \hhline{- - - -}
& $\Lambda_b^0 (udb)$ & $\pi^0, \,\pi^+ \, \pi^-$ & 5484, 5340 \\ 
& $\Sigma_b^+ (uub) $ & $\pi^0 \pi^+ $ & 5536 \\
& $\Sigma_b^- (ddb)$ &  $\pi^0 \pi^- $ & 5541 \\
 & $\Sigma_b^0 (udb)$ & $\pi^0, \,\pi^+ \pi^-$ & \emph{unknown} \\
& $\Xi_b^0 (usb)$ & \emph{none} & \emph{n.a}  \\
 & $\Xi_b^- (dsb)$ & $\pi^- \bar{K}^0 ,\, \pi^0 K^-$ &  5160, 5168\\
$\mathcal{O}_{bud}$ & $\Xi^0_{bb} (ubb)$ & $\pi^0 \bar{B}_d^0\,, \pi^+ B^-$ &  \emph{unknown}\\ 
$b \rightarrow \bar{u} \bar{d} \psiBbar $& $\Xi_{bb}^- (dbb)$ & $\pi^- B^0_d\,, \pi^0 B^-$ & \emph{unknown}\\ 
& $\Xi^+_{cb} (ucb)$ & $\pi^0 D^+, \, \pi^+ D^0$ & \emph{unknown} \\ 
& $\Xi^0_{cb} (dcb)$ & $\pi^0 D^0, \, \pi^- D^+$ & \emph{unknown}\\ 
& $\Omega_b^- (ssb)$ & $K^- \bar{K}^0$ & 5055 \\
& $\Omega_{cb}^0 (scb)$ & $\bar{K}^0 D^0 \,, K^+ D^-$ & \emph{unknown}  \\
& $\Omega_{bb}^- (sbb)$ & $K^- \bar{B}_d^0 \,, \bar{K}^0 B^-$ & \emph{unknown} \\
& $\Omega^+_{ccb} (ccb)$ & $D^0 D^+$ & \emph{unknown} \\
& $\Omega^0_{cbb} (cbb)$ & $D^+ B^- ,\, D^0 \bar{B}_d^0$ & \emph{unknown} \\ \hhline{- - - -}%%%%
& $\Lambda_b^0 (udb)$ & $\bar{D}^0 \pi^0,\, \pi^+ D^-$ & 3620, 3610 \\ 
& $\Sigma_b^+ (uub) $ & $\bar{D}^0 \pi^+$ & 3811 \\
& $\Sigma_b^- (ddb)$ &  $D^- \pi^0$ & 3806  \\
 & $\Sigma_b^0 (udb)$ & $\bar{D}^0 \pi^0 ,\, \pi^+ D^-$ & \emph{unknown} \\
& $\Xi_b^0 (usb)$ &  $\bar{D}^0 \bar{K}^0 ,\, \pi^+ D_s^-$ & 3430, 3684  \\
 & $\Xi_b^- (dsb)$ & $D^- \bar{K}^0,\, \pi^0 D_s^-$ & 3430, 3694 \\
$\mathcal{O}_{bcd}$ & $\Xi^0_{bb} (ubb)$ & $\bar{D}^0 \bar{B}_d^0 ,\, \pi^+ B_c^-$ &  \emph{unknown}\\ 
$b \rightarrow \bar{c} \bar{d} \psiBbar $& $\Xi_{bb}^- (dbb)$ & $D^- B_d^0,\,\pi^0 B_c^-$ & \emph{unknown}\\ 
& $\Xi^+_{cb} (ucb)$ & $\bar{D}^0 D^+,\, \pi^+ \eta_c$ & \emph{unknown} \\ 
& $\Xi^0_{cb} (dcb)$ & $D^- D^+,\, \pi^0 \eta_c$ & \emph{unknown}\\ 
& $\Omega_b^- (ssb)$ & $D_s^- \bar{K}^0$ & 3580 \\
& $\Omega_{cb}^0 (scb)$ & $D_s^- D^+,\, \bar{K}^0 \eta_c$ & \emph{unknown}  \\
& $\Omega_{bb}^- (sbb)$ & $D_s^- \bar{B}_d^0,\, \bar{K}^0 B_c^-$ & \emph{unknown} \\
& $\Omega^+_{ccb} (ccb)$ & $\eta_c D^+$ & \emph{unknown} \\
& $\Omega^0_{cbb} (cbb)$ & $\eta_c B_d^0,\, D^+ B_c^-$ & \emph{unknown} \\ \hhline{- - - -}
\end{tabular}
& 
\begin{tabular}{ |c || c | c | c  |}
\hhline{- - - -}
    Operator 			&  \,\, Initial \,\,  &  Final 	&     \,\,\, $\Delta M$     \,\,\,  \\     
    and Decay   			&  Baryon &  Mesons				& (MeV)       \\ \hhline{- - - -}
& $\Lambda_b^0 (udb)$ & $\pi^9 K^0, \, K^+ \pi^-$ & 4986 \\ 
& $\Sigma_b^+ (uub) $ & $ \pi^0 K^+ $ & 5172 \\
& $\Sigma_b^- (ddb)$ &  $\pi^- K^0$ & 5181 \\
 & $\Sigma_b^0 (udb)$ & $K^0 \pi^0 \,, K^+ \pi^-$ & \emph{unknown} \\
& $\Xi_b^0 (usb)$ &  $\eta', K^+K^-$ & 4834, 4804  \\
 & $\Xi_b^- (dsb)$ & $\pi^- \eta',\, K^0 K^-$ & 4699, 4805  \\
$\mathcal{O}_{bus}$ & $\Xi^0_{bb} (ubb)$ & $\pi^0 B_s^0,\, K^+ B^-$ &  \emph{unknown}\\ 
$b \rightarrow \bar{u} \bar{s} \psiBbar $& $\Xi_{bb}^- (dbb)$ & $\pi^- \bar{B}_s^0 ,\, K^0 B^-$ & \emph{unknown}\\ 
& $\Xi^+_{cb} (ucb)$ & $\pi^0 D_s^+, \, K^+ D^0$ & \emph{unknown} \\ 
& $\Xi^0_{cb} (dcb)$ & $\pi^- D_s^+,\, K^0 D^0$ & \emph{unknown}\\ 
& $\Omega_b^- (ssb)$ & $\eta K^-$ &  5005 \\
& $\Omega_{cb}^0 (scb)$ & $K^- D_s^+,\, \eta D^0$ & \emph{unknown}  \\
& $\Omega_{bb}^- (sbb)$ & $K^- \bar{B}_s^0,\, \eta B^-$ & \emph{unknown} \\
& $\Omega^+_{ccb} (ccb)$ & $D^0 D_s^+$ & \emph{unknown} \\
& $\Omega^0_{cbb} (cbb)$ & $D^0 \bar{B}_s^0 \,, D_s^+ B^-$ & \emph{unknown} \\ \hhline{- - - -}
%%%%
& $\Lambda_b^0 (udb)$ & $\bar{D}^0 K^0, K^+ D^-$ & 3257, 3256 \\ 
& $\Sigma_b^+ (uub) $ & $\bar{D}^0 K^+$ & 3452 \\
& $\Sigma_b^- (ddb)$ &  $D^- D^0$ & 2082  \\
 & $\Sigma_b^0 (udb)$ & $\bar{D}^0  K^0 ,\, K^+ D^-$ & \emph{unknown} \\
& $\Xi_b^0 (usb)$ &  $\bar{D}^0 \bar{B}_s^0$ & 1351 \\
 & $\Xi_b^- (dsb)$ & $D^- \eta, \, K^0 D_s^-$ &3386, 3331  \\
$\mathcal{O}_{bcs}$ & $\Xi^0_{bb} (ubb)$ & $\bar{D}^0 \bar{B}_s^0,\, K^+ \bar{B}_s^0$ &  \emph{unknown}\\ 
$b \rightarrow \bar{c} \bar{s} \psiBbar $& $\Xi_{bb}^- (dbb)$ & $D^- \bar{B}_s^0,\, K^0 B_c^-$ & \emph{unknown}\\ 
& $\Xi^+_{cb} (ucb)$ & $\bar{D}^0 D_s^+,\, K^+ \eta_c$ & \emph{unknown} \\ 
& $\Xi^0_{cb} (dcb)$ & $D^- D_s^+, \, K^0 \eta_c$ & \emph{unknown}\\ 
& $\Omega_b^- (ssb)$ & $\eta D_s^-$ & 3531  \\
& $\Omega_{cb}^0 (scb)$ & $D_s^+ D_s^- ,\, \eta \eta_c $ & \emph{unknown}  \\
& $\Omega_{bb}^- (sbb)$ & $D_s^- \bar{B}_s^0,
\,\eta B_c^-$ & \emph{unknown} \\
& $\Omega^+_{ccb} (ccb)$ & $\eta_c D_s^+$ & \emph{unknown} \\
& $\Omega^0_{cbb} (cbb)$ & $\eta_c \bar{B}_s^0,\, D_s^+ B_c^-$ & \emph{unknown} \\\hhline{- - - -}
%%%
\end{tabular}
\end{tabular}
\caption{Comprehensive summary of all possible heavy baryon decays into mesons and missing energy mediated by $\mathcal{Y}$. }
\label{tab:bBaryonDecays}
\end{table*}

%%%%%%%
\section{The Dark Sector}
\label{sec:darksector}
%%%%%%%%%%%%%%%%%%%

In $\dcp$ $\M$-Mesogenesis, the dark matter of the Universe consists of two components (stabilized kinematically and by a $\mathbb{Z}_2$ symmetry), namely $\xi$ and the symmetric component of the scalar baryon $\phiB + \phiB^*$. The dark matter production (and parameter space) in $\dcp$ $\M$-Mesogenesis is analogous to neutral $B$-Mesogenesis \cite{Elor:2018twp} and $B_c^+$-Mesogenesis \cite{Elahi:2021jia} mechanisms, as we now demonstrate. 

Without loss of generality, we require that $\psiB^1 \rightarrow  \phiB \xi$ be the dominant $\psiB^1$ decay mode and for simplicity assume $y_d^{\alpha \neq 1} = 0$. 
Since $\psiB^1$ decays rapidly (in comparison to the Hubble expansion) at MeV scales, using Eq.~\eqref{eq:psiBBE}, we arrive at the following Boltzamnn equations for the evolution of the (baryon number zero) fermionic dark matter component:  
\bea
 \frac{d n_\xi}{dt} + 3 H n_\xi &=&  2 \, \Gamma_\Phi \text{Br}_\Phi^{\M} \text{Br}_{\mathcal{M}}  n_\Phi \\ \nonumber
 && \qquad - \langle \sigma v \rangle_\xi \left( n_\xi^2 - n_{\xi , \,\rm eq}^2  \right) \,,
 \label{eq:DMxi}
\eea
and the scalar baryon dark matter component.  
  \bea
 &&\frac{d (n_{\phiB} + n_{\phiBbar})}{dt} + 3 H (n_{\phiB} + n_{\phiBbar}) =  \\ \nonumber
 &&2\, \Gamma_\Phi \text{Br}_\Phi^{\M}  \text{Br}_{\M}  n_\Phi - 2 \langle \sigma v \rangle_{\phiB} \left( n_{\phiB} n_{\phiBbar} -  n_{\phiB}^{\rm eq} n_{\phiBbar}^{\rm eq}\right) .
 \label{eq:DMphi}
\eea 
The above equations also include possible dark sector scattering contributions, with thermally averaged cross section denoted by $\langle \sigma v \rangle_d$. In addition to the $\phiB-\xi$ co-annihilation, such interaction control a dark sector freeze out which fixes the dark matter abundances such that $\Omega_\xi h^2 + \Omega_{\phiB + \phiBbar} h^2 = 0.1$ \cite{Elor:2018twp}. Pragmatically, without a specific dark sector model, we simply float the value of $\langle \sigma v \rangle_d$ to generate the correct dark matter abundance for any point in parameter space that generates the BAU in exact analog with the results of  \cite{Elor:2018twp,Elahi:2021jia}. For an example UV model  see \cite{Alonso-Alvarez:2019fym} in which the dark matter states were identified with a right handed neutrino super-multiplet and the associated phenomenology and dark matter abundance was computed.

Given a dark matter component carrying Standard Model baryon number, $\rho_{DM} \sim 5 \rho_{\mathcal{B}}$ fixes the relative ratios of $\phiB$ and $\xi$ in the dark matter halo today. Specifically, $\rho_\xi / \rho_{\phiB} = 5 m_p/m_{\phiB} - 1$ with $\rho_{\rm tot} = \rho_{\phiB} + \rho_\xi = 0.5\, \text{GeV}/\text{cm}^3$. 
To probe the dark sector, the dark matter induced nucleon decay search at neutrino experiments presented in \cite{Berger:2023ccd} would also serve as tests of $\dcp$ Mesogenesis.  

\begin{figure}[t!]
\centering
\includegraphics[width=0.45\textwidth]{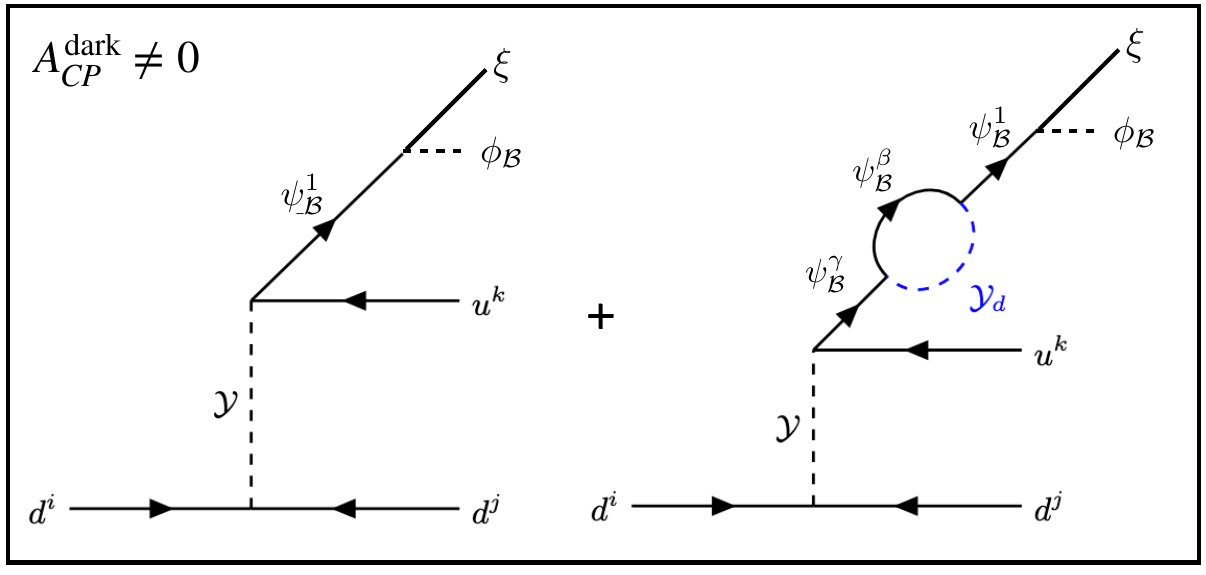}
\vspace{-.2cm}
\caption{Diagrams whose interferes leads to CP violation in meson decays when of-shell $\psiB$s decay to $\phiB \, \xi$.}
\label{fig:psi-production}
\end{figure}

We now discuss the dark sector ingredients required to generate CP violation $A_{CP}^{dark}\neq 0$ (specific UV models which generate order one $A_{CP}^{dark}$ is the subject of upcoming work \cite{utfolks}). One possible way to generate  dark sector CP violation is take $\psiB^{\alpha}$ to belong to a flavor triplet and introduce a second (GeV scale) dark mediator $\Y_d$ which is a scalar under the Standard Model gauge groups and does not carry baryon number.  Table~\ref{tab:newparticles} summarizes the new particle content of this toy model. In addition to Eq.~\eqref{eq:L23}, the model now includes dark sector Lagrangian interactions:
\begin{subequations}
    \begin{align}
& \mathcal{L}_{dark}^{int} = - \sum_{\alpha,\beta} \lambda_{\alpha \beta} \Y_d \psiBbar^\alpha \psiB^\beta -\sum_\alpha y_d^\alpha  \psiB^\alpha \phiB \xi  + \text{h.c}  \\
& \mathcal{L}_{mass}^{\psi}  = - \sum_{\alpha \beta} M_{\alpha \beta} \psiBbar^\alpha \psiB^\beta+ \text{h.c} \,,
\label{eq:L_mass}
    \end{align}
\end{subequations}
Given this construction, physical dark CP violating phases exist when there are more than two dark sector fermion flavors \cite{Nir:2007xn} (just like in the Standard Model).  
%Note that baryon number forbids mass mixing between $\psiB$ and SM quarks. 
%$\mathcal{L}_{dark}$ allows the lightest flavor of $\psiB$ to decay to the DM particles $\psiB^1 \rightarrow \phiB \xi$ as previously mentioned.
$A_{CP}^{\rm dark} \neq 0$ in the decay $\M \rightarrow \mathcal{B} + \text{MET}$ arises from the interference between the tree level and 1-loop decays involving these dark CP violating phases. This can be illustrated in the usual way: writing the decay rate as $\Gamma \propto \int d\Pi_3 |\mathcal{A}_t +  \mathcal{A}_l|^2$ with the scattering amplitude, $\mathcal{A}$, separated into tree and 1-loop level contributions and where $d\Pi_3$ is the three-body phase space. The CP violation observable is then goes as $A_{CP}^{\rm dark} \propto \text{Im} [\A_t \A_l^*] /|\A_t|^2$. Figure \ref{fig:psi-production} shows possible parton level  (meson to baryon form factors essentially ``cancel out") diagrams contributing to  $A_{CP}^{\rm dark}$. In such process, CP violation is can enhanced when the internal propagator goes on-shell  \cite{Pilaftsis_2004,PhysRevD.56.5431,PhysRevD.56.5431} and as such sizable (close to one) CP violation may arise. Indeed the calculation parallels that of leptogenesis constructions \cite{Garny_2013} as well as certain twin-sector models \cite{Kilic:2021zqu}. Note that lighter mediator, $\Y_d$, was introduced solely for generating $A_{CP}^{dark} \neq 0$. Possible simpler models, along with their UV completions and associated phenomenology, which can generate order one dark CP violation is the subject of study in upcoming work \cite{utfolks}. 

\begin{table}[t!]
\renewcommand{\arraystretch}{1.25}
  \setlength{\arrayrulewidth}{.25mm}
\centering
\small
\setlength{\tabcolsep}{0.18 em}
\begin{tabular}{|c || c | c| c | c | c  | c |}
\hhline{- - - - - - -}
Field & Spin & SU$(3)_c$ & $U(1)_Y$ & $Q_{\mathcal{B}}$ & $\mathbb{Z}_2$ & Mass \\ \hhline{- - - - - - -}
$\Phi$ & 0 & \textbf{1} & 0 & 0 &+1 & $2m_{\M}$ -100GeV \\
\hhline{- - - - - - -}
$\Y$ & 0 & \textbf{3} & 2/3 & -2/3 & +1 & $\mathcal{O} (\text{TeV})$ \\ \hhline{- - - - - - -}
$\Y_d$ & 0 & \textbf{1} & 0 & 0 & +1 & $\mathcal{O}(\text{GeV})$ \\ \hhline{- - - - - - -}
$\psiB^{l}$ & 1/2 & \textbf{1} & 0 & -1 & +1 & $\mathcal{O}(\text{GeV})$  \\ \hhline{- - - - - - -}
$\phiB$ & 0 & \textbf{1}& 0 & -1 & -1 & $\mathcal{O}(\text{GeV})$ \\ \hhline{- - - - - - -}
$\xi$ & 1/2 & \textbf{1} & 0 & 0 & -1 & $\lesssim \mathcal{O}(\text{GeV})$ \\ \hhline{- - - - - - -}
\end{tabular}
\caption{The BSM particle content and their quantum numbers of a minimal toy model realizing  $\dcp$ Mesogenesis. A $\mathcal{Z}_2$ symmetry is assumed to stabilize the dark matter particles $\phiB$ and $\xi$. Here the dark fermion flavor index $\l \geq 2$.} 
\label{tab:newparticles}
\end{table}

\begin{figure}[t!]
\centering
\includegraphics[width=0.3\textwidth]{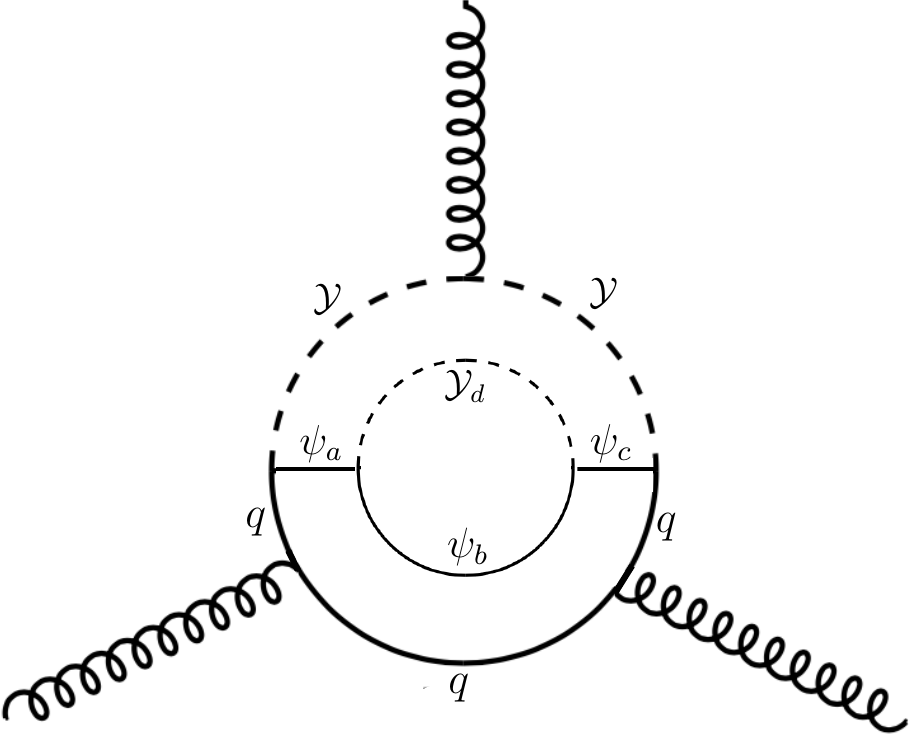}
\vspace{-.2cm}
\caption{A Feynman diagram involving the dark sector states contributing to the CP-violating three gluon operator the Standard Model. A similar diagram where two gluons attach to the $\mathcal{Y}$ line and one to the quark line is not shown.}
\label{fig:EDM}
\end{figure}

Flavor constraints are identical to past Mesogenesis proposals involving dark sector baryons --- see \cite{Alonso-Alvarez:2021qfd}. New phenomenology associated with $\dcp$ Mesogenesis will be in the form of 2-loop contributions to hadronic electron dipole moments as shown in Figure \ref{fig:EDM}. The Standard Model prediction for the neutron electric dipole moments (nEDM) arising from the CKM phase is: $d_n^{\rm SM} \simeq 10^{-31} e\, \text{cm}$ \cite{dar2000neutronedmsm}. This lies several orders of magnitude below the current experimental upper bound: $d_n = (0.0\pm 1.1) \times 10^{-26} e\, \text{cm}$ \cite{Abel:2020pzs} --- leaving room for new physics contributions beyond the Standard Model to CP violation.  The nEDM contribution from Figure \ref{fig:EDM} can be estimated by mapping onto the Weinberg operator for gluon fields~\cite{Abe:2017sam}: $\mathcal{O}_{W} = \frac{1}{3} C_G f^{abc} \epsilon^{\mu \nu \rho \sigma} G_{\mu \lambda}^a G_\nu^{a \lambda} G_{\rho \sigma}^c$. Then $\Delta d_n = \pm e \Lambda_{\text{nEDM}} g_s^3 C_G$,  where $\Lambda_{\text{nEDM}}  = 10-30$ MeV. Roughly,  $C_G \propto \frac{1}{M_{\mathcal{Y}}^4} \frac{\Lambda_{QCD}^2}{(16 \pi^2)^3}  g_s^3$ times the imaginary part of the product of vertex couplings in Figure \ref{fig:EDM}. Given that $M_{\mathcal{Y}}$ is constrained to be at the TeV scale by collider searches \cite{Alonso-Alvarez:2021qfd}, even for order one phases, such diagrams would not be excluded by current nEDM measurements. However, future improvements on measurements of nEDMs \cite{pignol2021searchneutronelectricdipole,Ayres_2021} could begin to probe such models. 
%%%%%%%%%%%%%%%%%%

%
\section{Low-Scale Leptogenesis}
\label{sec:lowscalelepto}
\begin{figure*}[t!]
\centering
\includegraphics[width=0.9\textwidth]{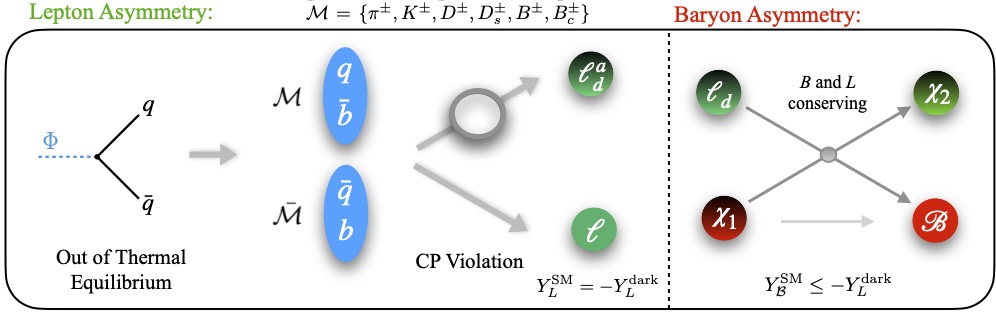}
\vspace{-.2cm}
\caption{Depiction of $\dcp$ Mesogenesis variant with dark leptons}
\label{fig:cartoonDarkLeptons}
\end{figure*}

A goal of this work is to comprehensively explore the parameter space of the Mesogenesis framework while leveraging dark sector CP violation. To this end we now briefly introduce \emph{$\dcp$ Mesogenesis with Dark Leptons}---a class of mechanisms where a lepton asymmetry is first generated through CP violating decays of charged mesons $\mathcal{M} = \left\{ \pi^\pm, K^\pm, D^\pm, D^\pm_s, B^\pm, B_c^\pm \right\}$ and anti-mesons to a visible and dark sector leptons:  $\mathcal{M}^\pm \rightarrow l + l_d$, where the Standard Model lepton $l$ may be an electron, muon, or possibly even tau for $B_c^\pm$ decays. The generated lepton asymmetry is then transferred into a baryon asymmetry through dark sector scatterings. This is analogous to Charged $D$ and $B$ Mesogenesis \cite{Elor:2020tkc,Elahi:2021jia} but is notably simpler as CP violation is possible without multi-step decays. Figure \ref{fig:cartoonDarkLeptons} illustrates the mechanism.

The CP violating meson decays can arise through an operator of the form: 
\bea
\mathcal{O} = \frac{1}{\Lambda^2} \sum_{a} \sum_{l = e, \mu, \tau} \left[ \bar{d} \Gamma^\mu u \right] \left[\bar{l}_d^a \Gamma_\mu l \right] + \text{h.c.}\,,
\eea 
where $\Gamma_\mu$ encodes the Lorentz structure. Physical dark sector phases can lead to sizable CP violation between meson and anti-meson decays with the introduction of multiple dark lepton flavors $l_d^a$ and lepton mass mixing: 
\bea
\mathcal{L}= \sum_{ab} M_{ab} \bar{l}_d^a  l_d^b + \sum_{ab} \lambda_{ab} S \bar{l}^a_d l^b_d\,,
\eea 
where $S$ is a light mediator.  

Production and decay of $\M^\pm$ mesons is tracked through:
\begin{subequations}
\begin{align}
& \frac{dn_{\M^+}}{dt} + 3 H n_{\M^+} = \Gamma_\Phi^{\M}n_\Phi - \Gamma_{tot} n_{\M^+}\,,  \\ 
&  \frac{dn_{\M^-}}{dt} + 3 H n_{\M^-} = \Gamma_\Phi^{\M}n_\Phi - \bar{\Gamma}_{tot} n_{\M^-} \,,
\end{align}
\end{subequations}
where $\Gamma_\Phi^{\M^+} = \Gamma_\Phi^{\M^-} \equiv \Gamma_\phi^\M$, since once again we assume that equal numbers of $\M^+$ and $\M^-$ mesons are produced in the decay of $\Phi$. For low temperatures and rapid decays, the produced mesons quickly decay leading to the following evolution of the dark lepton number density: 
\begin{subequations}
\begin{align}
& \frac{d n_{l_d}}{dt} + 3 H n_{l_d} =  \simeq \frac{\Gamma(\M^+ \rightarrow l^+ \, l_d)}{\Gamma_{tot}} \Gamma_\Phi^\M n_\Phi \,,
\end{align}
\end{subequations}
Following the derivations in \cite{Elor:2020tkc,Elahi:2021jia}, the lepton asymmetry, $Y_L^{dark} \equiv (n_{l_d} - \bar{n}_{\bar{l}_d}) / s = - Y_L^{SM}$, is generated through the production of dark leptons $l_d$ and anti-leptons $\bar{l}_d$, and the lepton asymmetry is governed by:
\bea
&& \frac{d (\bar{n}_{\bar{l}_d}-n_{l_d} )}{dt} + 3 H (\bar{n}_{\bar{l}_d}-n_{l_d}) \\ \nonumber
&& = \frac{1}{\Gamma_{tot}} \left[ \Gamma(\M^+ \rightarrow l_i^+\, l_d) -  \Gamma(\M^- \rightarrow l_i^- \, \bar{l}_d)\right] \Gamma_\Phi^\M n_\Phi \\ \nonumber
&& = 2  \text{Br}_{\M \rightarrow l_i l_d} A_{CP}^{dark, l} \Gamma_\Phi^\M n_\Phi\,.
\label{eq:DMesoYL}
\eea
Here we have introduced the new dark sector CP violating observable: 
\bea
A_{CP}^{dark, l}  \equiv  \frac{\Gamma(\M^+ \rightarrow l^+ \, l_d) -  \Gamma(\M^- \rightarrow l^- \, \bar{l}_d)}{\Gamma(\M^+ \rightarrow l^+ \, l_d) +  \Gamma(\M^- \rightarrow l^- \, \bar{l}_d)}\,.
\eea
Numerically solving these equations is described in Eq.~\eqref{eq:DMesoYL} \cite{Elor:2020tkc,Elahi:2021jia}, and arrive at the following lepton asymmerty:
\bea
\hspace{-0.2in}
\frac{Y_L^{SM}}{Y_{\mathcal{B}}^{obs}} \simeq
\sum_{i = e,\mu, \tau} \frac{\text{Br}_{\M^\pm \rightarrow l_i l_d}}{3 \times 10^{-6}} \frac{A_{CP}^{dark}}{10^{-2}} \frac{T_R}{20 \, \text{MeV}} \frac{10 \, \text{GeV}}{m_\Phi}\,.
\eea
Leading constraints on $\text{Br}_{\M^\pm \rightarrow l_i l_d}$, the branching fraction of $\M^\pm \rightarrow l_i l_d$ decays, comes from re-casting neutrino peak searches (see \cite{Elor:2020tkc,Elahi:2021jia}). Such constraints depend on the missing energy and as such only constrain a subset of kinematically allowed $m_{l_d}$ masses. Thus the mechanism is currently only weekly, at best, constrained. 

This variant of $\dcp$ Mesogenesis opens up the parameter space of possible MeV scale leptogenesis first introduced in \cite{Elor:2020tkc}. Upcoming work \cite{LowScaleLepto} will explore a full model and associated phenomenology. 

\section{Roadmap}
\label{sec:dis}
%%%%%%%%%%%%%%%%%
With the introduction of $\dcp$ $\M$-Mesogenesis, we approach a complete map of the theory space of Mesogenesis mechanisms. This serves an important role in enabling the discovery or ultimately the complete exclusion of the Mesogenesis framework.  While we have only presented a minimal model, our conclusions should apply to more elaborate models of this class and will be explored in upcoming work \cite{utfolks,LowScaleLepto}. 

Current experimental searches for Mesogenesis (at Belle-II and LHCb) focus on measuring $\text{Br} \left( B^0 \rightarrow \mathcal{B} + \text{MET}\right)$ down the $10^{-6}$ to $10^{-5}$ level, below which neutral $B^0$ Mesogenesis \cite{Elor:2018twp} is excluded. Therefore a \emph{key message} of this work to our experimental friends is to search for all the exotic decays listed in Table \ref{tab:decays} down to $10^{-8}$ sensitivity in branching fraction---only below this value will the Mesogenesis framework (with dark baryons) becomes disfavored. In contrast, a detection with a branching fraction the range $10^{-8} \lesssim \text{Br} \lesssim 10^{-5}$ would be  \emph{strong evidence for} $\dcp$ Mesogenesis, which, in conjunction with the reconstruction of $A_{CP}^{dark}$ from $B$-factories searches, indirect signatures at LHCb, and possible future EDM measurements, would allow for full testability of $\dcp$ Mesogenesis.   

We now return to the question of whether Mesogenesis (in any variant) can ever be fully excluded? Should the decays listed in Table \ref{tab:decays} all be probed down to $10^{-8}$ sensitivity in branching fraction, Mesogenesis with dark baryons will be excluded up to one caveat: Mesogenesis augmented with a dark sector phase transition which \emph{morphs} the mass of one of the new particles in Table \ref{tab:newparticles} after the BAU is generated \cite{Elor:2024cea,technion}. This effectively decouples the rate $\Gamma \left( \M \rightarrow \mathcal{B} + \text{MET}\right)$ at colliders today from its value in the early Universe. Upcoming work, \cite{technion}, will investigate the degree to which tuning is required in the requisite late-time dark phase to successfully generate the BAU in this fashion.  A null discovery of $\M \rightarrow \mathcal{B} + \text{MET}$,  coupled with theory work indicating that the only caveat requires a highly tuned phase transition, would be very strong motivation for excluding Mesogenesis variants involving dark sector baryons. After this the only remaining possibility will be $B^+$ and $D^+$ Mesogenesis and the variants of $\dcp$ Mesogenesis with dark leptons introduced in Section \ref{sec:lowscalelepto}---where CP violating meson decays first produce a lepton asymmetry which is later transferred to a baryon asymmetry through dark sector scatterings. Such variants, while more complex, are still testable at current and future experiments. However signals tend to be more model dependent and will be further explored in future work \cite{LowScaleLepto}. 

In conclusion, this work serves as a roadmap to guide experimentalist and theorists alike towards for the complete exclusion (or discovery) of the Mesogenesis framework.

\begin{acknowledgments}
I thank Yotam Soreq, Can Kilic, Sanjay Mathai for useful discussions about dark sector CP violation.  I thank the hospitality of CERN (where this project was conceived) and further thank CERN for support through the theory visitor program. I also thank Lawrence Berkeley National Laboratory for providing funding for a trip to KEK. Completion of this work was further enabled through the support of Stone Aerospace.  
\end{acknowledgments}

\bibliography{Refs}
\bibliographystyle{utphys}

\end{document}